# Ultrafast Exciton-Polariton Transport and Relaxation in Halide Perovskite

Yuexing Xia[1,2], Yin Liang[3], Zhiyong Zhang[1], Tian Lan[3], Xiaotian Bao[1], Yuyang Zhang[3], Xin Zeng[1], Yiyang Gong[1,2], Shuai Yue[1,2], Wenna Du[1,2], Jianhui Fu[1,2], Rui Su[4], Stefan Schumacher[5], Xuekai Ma[5]*, Qing Zhang[3]*, Xinfeng Liu[1,2]*

[1]CAS Key Laboratory of Standardization and Measurement for Nanotechnology, National Center for Nanoscience and Technology, 100190 Beijing, China.

[2]University of Chinese Academy of Sciences, 100049 Beijing, China.

[3]School of Materials Science and Engineering, Peking University, 100871 Beijing, China.

[4]Division of Physics and Applied Physics, School of Physical and Mathematical Sciences, Nanyang Technological University, 637371 Singapore, Singapore.

[5]Department of Physics and Center for Optoelectronics and Photonics Paderborn (CeOPP), Paderborn University, 33098 Paderborn, Germany.

*Corresponding authors: liuxf@nanoctr.cn, q_zhang@pku.edu.cn, xuekai.ma@gmail.com

## Abstract

Halide perovskites offer a great platform for room-temperature exciton-polaritons (EPs) due to their strong oscillator strength and large exciton binding energy, promising applications in next-generation photonic and polaritonic devices. Efficient manipulation of EP transport and relaxation is critical for device performance, yet their spatiotemporal dynamics across different in-plane momenta ($k_{//}$) remain poorly understood due to limitations in experimental access. In this work, we employ energy-resolved transient reflectance microscopy (TRM) combined with the dispersion relation of EPs to achieve high-resolution imaging of EP transport at specific $k_{//}$. This approach directly reveals the quasi-ballistic transport and ultrafast relaxation of EPs in different $k_{//}$ regions, showcasing diffusion as fast as ~490 cm$^2$/s and a relaxation time of ~95.1 fs. Furthermore, by tuning the detuning parameter, we manipulate the ballistic transport group velocity and relaxation time of EPs across varying $k_{//}$. Our results reveal key insights into the dynamics of EP transport and relaxation, providing valuable guidance for the design and optimization of polaritonic devices.

# Main

Exciton-polaritons (EPs) are hybrid light-matter quasiparticles arising from strong exciton-photon coupling, endowing them with the photons' high coherence and low effective mass, together with the excitons' strong nonlinearity[1,2]. Owing to this hybrid character, EPs constitute a versatile platform for exploring fundamental physics − from non-equilibrium quantum phenomena to solid-state Bose-Einstein condensation[3-6] − as well as for enabling disruptive technologies such as ultrafast logic gates[7,8], low-threshold lasers[9-12], and quantum simulators[13-15]. However, the practical realization of such devices has a critical dependence on two factors: efficient in-plane coherent transport[16-19] and rapid dynamic response[20-23]. These key metrics are governed by the spatiotemporal dynamics of the lower polariton branch (LPB), which themselves hinge on the intricate interplay of scattering[24,25], relaxation[26-28], and spatial propagation[29-31]. Consequently, the direct, time-resolved measurement of these dynamics remains a central limitation, especially in emergent room-temperature exciton-polariton systems like halide perovskites[32-38].

In this work, we develop an in-plane momenta ($k_{//}$) resolved pump-probe microscopy technique that enables the direct visualization of the spatiotemporal dynamics of ultrafast in-plane transport and relaxation process of EPs. Using this platform, we unravel the spatiotemporal dynamics of EPs in 2D halide perovskite, and observe the distinct in-plane transport and relaxation of EPs intermediate and near branch minimum ($k_{//} \approx 0$). At finite $k_{//}$, we observe a fast quasi-ballistic coherent polariton transport following by a transient contraction of transport towards exciton-like diffusion; at near-zero-$k_{//}$, we observe a slow ballistic coherent polariton transport following by the dephasing process. Furthermore, we quantify the momentum-dependent ballistic velocity and ballistic relaxation time governed by the photonic fraction. This understanding then allows us to actively engineer the spatiotemporal dynamics through precise control of the energy detuning. Our findings offer a reliable platform to investigate the intricate relaxation-transport interplay and reveal the microscopic mechanisms governing the ultrafast ballistic transport and relaxation of exciton-polaritons.

## Results

**Spatiotemporal dynamics of EPs at a specific in-plane momentum**

**Fig. 1a** shows the schematics of as-designed in-plane momentum ($k_{//}$) resolved transient reflectance microscopy (TRM), based on a pump-probe technique, to explore spatiotemporal dynamics of EPs. The TRM, directly quantifying the 2D image of photo-excitation induced dielectric response in time domain, offers superior temporal resolution compared to time-resolved photoluminescence image[39,40]. Herein, a focused non-resonant pump pulse (400 nm, 100 fs) rapidly populates the exciton reservoir, which subsequently thermalizes into the LPB through phonon scattering. A narrow-band probe beam, energy-matched with LPB (such as 532 nm), is slightly divergent to fully illuminate the back focal plane of the objective lens, forming a ~5 μm spot on the sample surface. The strong light-matter coupling in semiconductor microcavity enforces that the polaritons are distributed along their dispersion relation. Consequently, each probe wavelength corresponds to a specific $k_{//}$ on the LPB, allowing $k_{//}$-selective probing of EP dynamics simply by tuning the probe wavelength. This configuration ensures complete coverage of the $k_{//}$ space while maintaining wide-field imaging capability.

We next applied this technique to a planar Fabry-Pérot (FP) microcavity incorporating 2D perovskite of butylammonium lead iodide (hereafter, BAPI, **Fig. S1**). The 2D perovskite provides strong exciton binding and oscillator strength, enabling robust EP formation at room temperature[32]. As shown in **Fig. 1b**, the angle-resolved reflectance spectroscopy revealed the EP dispersion relation and a substantial Rabi splitting energy of 223.8 meV (a detailed strong coupling analysis is presented in **Fig. S4-5**). The selection of $k_{//}$ was directly based on this established dispersion relation; for example, a probe wavelength of 535 nm corresponds to the LPB at $k_{//} = 5.2$ μm$^{-1}$. Next, to ensure a linear relationship between the transient reflectance intensity (ΔR/R) and the polariton density ($N$), we determined the suitable pump fluence range via power-dependent transient reflectance measurements (**Fig. S6-7**). Within this linear regime, TRM was performed to obtain 2D maps of the polariton population distributions at various time delays (**Fig. 1c**). For the quantitative analysis we fit the each TRM snapshot with a 2D Gaussian function, $N(x,y,t) = N_0 exp\{-[x^2/2\sigma_x^2(t) + y^2/2\sigma_y^2(t)]\}$, where $\sigma_x^2(t)$ and $\sigma_y^2(t)$ denote the variances along two orthogonal in-plane directions. Given that the distributions are nearly circular, indicating isotropic in-plane transport, we define a single variance $\sigma^2(t) = \sigma_x^2(t) = \sigma_y^2(t)$ to characterize the polariton distribution. As shown in the **Fig. 1d**, the good agreement between experimental image and fitting profile suggest the applicability of the Gaussian model. Then, we extracted the $\sigma^2(t)$ from each snapshot and plotted it as a function of time delay. As shown in **Fig. 1e**, the temporal evolution of the variance within 36 ps, revealing three stages of dynamical behavior: (**i**) rapid expansion from the initial spot size before 0.6 ps, (**ii**) sharp contraction back to near the initial size between 0.6 and 1.12 ps, and (**iii**) subsequent slow diffusion beyond 1.12 ps.

To further quantify these three transport regimes, we extract the diffusivity *D* and exponent *α* by fitting the mean-square displacement (MSD) to the power law $\text{MSD} = 2Dt^\alpha$.[40] As a reference for exciton transport, we measured the diffusion of a bare BAPI flake under identical conditions, yielding a normal diffusive regime (*α* = 1) with *D* = 0.08 cm²/s (gray curve in **Fig. 1e**), corresponding to pure exciton diffusion. While, for the halide perovskite microcavities, at stage (**i**), we observed a superdiffusive process with *α* = 1.55; since the hot-carrier effects were ruled out, the 1 < *α* < 2 indicates quasi-ballistic transport, which likely results from the combined contribution of coherent polariton ballistic transport and excitonic normal diffusion. The diffusivity, *D* = 490 cm²/s, is 6000 times larger than that of the pure exciton normal diffusion, highlighting the remarkably enhanced in-plane transport capability of EPs in this system. The transport in stage (**iii**) returns to a normal diffusive regime with *α* ≈ 1. The diffusivity of *D* = 0.12 cm²/s is close to that of the pure exciton, indicating that transport is primarily governed by exciton-like particles in this stage. Notably, the contraction behavior observed in stage (**ii**) manifests as an apparent negative diffusivity, suggesting a transient localization of the population. The underlying mechanism will be further analyzed in the following sections.

**Spatiotemporal dynamics of $k_{//}$-dependent EPs**

To elucidate the relaxation pathways and transport behaviors of EPs on the LPB, we performed TRM imaging at both finite- and zero- $k_{//}$. This approach enabled a direct comparison of the spatiotemporal dynamics of EPs at intermediate LPB positions and near the band bottom. Specifically, we selected $k_{//}$ = 3.8 μm$^{-1}$ (**Fig. 2a**) and $k_{//}$ = 0 μm$^{-1}$ (**Fig. 2b**) recording TRM images within the first 10 ps, respectively. To facilitate visualization of their spatiotemporal evolution of and EP distribution, these images were normalized along the one spatial axis. At $k_{//}$ = 3.8 μm$^{-1}$, the EP population initially expanded and subsequently underwent a clear contraction. This observation is consistent with our earlier findings at finite-$k_{//}$, reflecting quasi-ballistic transport followed by rapid dissipation. In contrast, no contraction was observed at $k_{//}$ = 0 μm$^{-1}$; Instead, the spatial distribution transitioned from an initial expansion to a steady linear diffusion regime. To quantify the observed evolution, we performed 2D Gaussian fitting on each image frame to extract the variance, and plotted its temporal evolution in **Fig. 2c**. The finite-$k_{//}$ population exhibited quasi-ballistic transport before 0.4 ps, followed by a sudden contraction of the variance. In contrast, the population at $k_{//}$ = 0 μm$^{-1}$ showed no such contraction; instead, its variance steadily increased after an initial quasi-ballistic regime, consistent with normal diffusion dominated by excitons. Notably, the quasi-ballistic regime for the $k_{//}$ = 0 μm$^{-1}$ population extended longer, close to 1 ps. At the end of the quasi-ballistic regime, the subsequent diffusion rates of both the finite- and zero-$k_{//}$ populations were similar, consistent with diffusion occurring in an exciton-like manner.

To validate the experimentally observed dynamics, we performed numerical simulations using a simplified Gross-Pitaevskii model (see Methods) that neglects complex scattering processes[21,41]. Our simulations yielded time-dependent distributions of both the exciton reservoir and polariton densities under the similar non-resonant excitation as in experiments. A Fourier transform was applied to the time evolutions to isolate their populations within specific $k_{//}$ intervals by utilizing a narrower filter (~5 nm) in the frequency domain. The resulting real spatial distribution for each $k_{//}$ range was then obtained via an inverse Fourier transform. **Fig. 3a** shows the normalized polariton distribution for $k_{//}$ = 2.5 μm$^{-1}$ up to 1.13 ps. Notably, the polariton density at this $k_{//}$ is observed to approach zero near 1.13 ps, indicating a rapid and near-complete depopulation of this state. Consequently, the distribution results for subsequent times are not presented. To better align with our experimental observations, we instead present the normalized exciton distribution after 2.64 ps. The density distribution in between (1.13 < $t$ < 2.64 ps) was interpolated by using an exponential decay function. It can be seen that the exciton distribution remains nearly unchanged relative to its initial state, which suggests that the exciton reservoir population is largely stable after the initial polariton decay. In contrast, at $k_{//}$ = 0 μm$^{-1}$ (**Fig. 3b**), the polariton density decays more gradually. This behavior reflects the lack of efficient relaxation channels and is consistent with the expected transition of polaritons from a coherent (ballistic) to an incoherent (diffusive) transport regime.

Subsequently, we computed the MSD evolution for polariton densities at $k_{//}$ = 2.5, 2.0, and 0 μm$^{-1}$, as shown in **Fig. 3c**. Notably, the MSD curves for $k_{//}$ = 2.5 and 2.0 μm$^{-1}$ were seamlessly joined to the corresponding exciton MSD via an exponential fit (represented by dashed lines). This approach effectively models the transition from polariton to exciton-like behavior as the system evolves. The theoretical trends derived from these MSD profiles show excellent agreement with our experimental observations, both qualitatively and in terms of the observed ballistic-to-diffusive transition.

Based on our experimental data and numerical calculations, we propose the following schematic interpretation (see **Fig. 3d**). At the initial moment, the polariton populations at $k_{//}$ > 0 and $k_{//}$ = 0 are essentially similar, although their exciton fractions may differ. On a short timescale (0-$t_1$), polaritons at $k_{//}$ > 0 exhibit rapid expansion, characteristic of quasi-ballistic transport of coherent polaritons. Concurrently, polariton-polariton and polariton-phonon scattering processes lead to dispersion-driven relaxation of the polariton population, resulting in a rapid decrease or even disappearance of the polariton signal at $k_{//}$ > 0 ($t_1$-$t_3$). At this moment, the overall distribution nearly returns to its initial size, corresponding to a small amount of excitons remaining at $k_{//}$ > 0. These excitons likely originate from the band edge or exciton states that are broadened by exciton-phonon scattering. In contrast, the polariton population at $k_{//}$ = 0 is sustained for a longer duration (0-$t_2$), owing to relaxation inflow from the finite-$k_{//}$ states. As polaritons undergo dephasing ($t_2$-$t_3$), the limited availability of further relaxation pathways at $k_{//}$ = 0 causes dephasing to primarily occur *in situ*. This results in the preservation of the particle distribution in the final images. Taken together, these observations collectively demonstrate an intimate link between the contraction of polariton distributions at finite-$k_{//}$ and their relaxation dynamics. Building upon this understanding, the subsequent section is dedicated to the quantitative evaluation of the polariton group velocity and relaxation time, achieved through an integrated analysis that encompasses both ballistic transport and distribution contraction processes.

**Manipulation of EP transport dynamics by detuning.**

The transport behavior of EPs is fundamentally dictated by their photonic fraction $C_{ph}$, which can be precisely controlled via the microcavity detuning ($\Delta$). To vary the strong coupling conditions, we modulated the cavity photon mode energy by adjusting the thickness of the BAPI layer. Initial guidance for the experimental design was provided by transfer matrix method calculations, which yielded the thickness-dependent reflection spectra (**Fig. S11**). To ensure a two-branch system and minimize potential interactions from higher-order polariton branches, the BAPI layer thickness was restricted to the range of 50-200 nm. By using the angle-resolved reflection spectra (**Fig. S12**), we successfully selected three distinct detunings with $\Delta$ = 60.3, -95.2 and -166.6 meV, and displayed the time evolution of the MSD in **Figs. S13**, **4a** and **4b**, respectively.

To quantitatively analyze the transport behavior of EPs, we employed a piecewise distribution model. In the first regime, the transport exhibits a ballistic behavior characterized by an exponential decay, while in the second regime, the distribution evolution is governed by exciton-dominated diffusion. The model can be expressed as follows:

$$\sigma(t) = \begin{cases} \sigma_0 + \dfrac{(v_g^{(b)} t)^2}{\sigma_0} exp(-t/\tau_{relax}), t \leq t_p \\ \sqrt{\sigma_0^2 + 2Dt}, t > t_p \end{cases} \quad (1)$$

Here, $v_g^{(b)}$ denotes the group velocity of the ballistic transport, $\tau_{relax}$ denotes the duration of the ballistic regime, and $t_p$ corresponds to the time at which the transport becomes predominated by diffusion. It is noteworthy that the ballistic transport reflects the existence of coherent polaritons, and its termination indicates that the polaritons have relaxed from the probe region. Therefore, the duration of the ballistic regime is taken as a measure of the polariton relaxation time. Based on this model, the $v_g^{(b)}$ and $\tau_{relax}$ of polaritons at different $k_{//}$ under various detuning conditions are extracted, as shown in **Figs. 4c** and **4d**. In the theoretical limit of an ideal case, the group velocity of polaritons can be directly derived from the extracted polariton dispersion (**Fig. S12**) using the relation $v_g = \partial\omega/\partial k$. The theoretical $v_g$ values shown in Fig. 4c, however, were scaled by a factor of 0.1 to better fit the experimental data (**Fig. 4c**). There are two main reasons to illustrate the significantly reduced group velocities observed in experiments. One is the local potentials, such as the unavoidable disorders on top of the samples as well as the optically induced ones, that hinder the outgoing propagation of the polaritons. The other is the finite polariton lifetime, which restricts the expansion of the spot, i.e., $\sigma(t)$, in picosecond scales and consequently decreases the calculated group velocities by using Eq. (1). Our analysis under various detuning conditions reveals that the ballistic group velocity increases with $k_{//}$. Moreover, a higher photonic fraction achieved by increasing the negative detuning energy further enhances the $v_g^{(b)}$ at large $k_{//}$, which is consistent with theoretical predictions. For instance, at $k_{//} = 6.2$ μm$^{-1}$, the $v_g^{(b)}$ is 0.54 μm/ps for $\Delta = 60.3$ meV; while it can be enhanced to 2.07 μm/ps for $\Delta = -166.6$ meV. The $\tau_{relax}$ decreases with increasing $k_{//}$, as high-energy polaritons are thermodynamically unstable and preferentially relax toward lower-energy states. Notably, $\tau_{relax}$ does not decrease monotonically with $k_{//}$; a slight rebound appears at higher $k_{//}$, which can be attributed to the polariton bottleneck effect. Furthermore, the enhanced photonic fraction under stronger negative detuning accelerates the relaxation through cavity photon decay and facilitates more efficient scattering and energy transfer with low-energy states. In our measurements, the relaxation time can be as short as approximately 95.1 fs. These findings provide a feasible strategy for manipulating EP dynamics and realizing ultrafast polaritonic responses.

## Conclusion

In this work, we probed the spatiotemporal dynamics of EPs in 2D perovskite microcavities by employing a home-built momentum-resolved TRM. At finite-$k_{//}$, we observed ultrafast coherent polariton ballistic transport within the initial several hundreds of femtoseconds at a rate approximately 6000 times ($D = 490$ cm$^2$/s) faster than exciton diffusion. In addition to this early-time transport, polaritons exhibit a contraction of their spatial distribution as a manifestation of relaxation into band-edge or exciton-broaden states with normal diffusivity behavior. As approaching zero-$k_{//}$ we observed a slower ballistic transport followed by a polariton dephasing processes. Moreover, we uncovered the momenta and detuning-dependent transport and relaxation dynamics, observing that the group velocity increases and the relaxation time decreases with $k_{//}$ under the same detuning, and that an increased photonic fraction further enhances the group velocity and shortens the relaxation time at a given $k_{//}$. The group velocity can be enhanced up to 2.07 μm/ps by tuning the cavity detuning, while the relaxation time can be as short as 95.1 fs. These findings offer direct experimental insight into the non-equilibrium relaxation mechanisms governing polariton transport. Furthermore, they also provide a practical strategy for manipulating polariton flow and achieving ultrafast polaritonic functionalities in room-temperature perovskite microcavities.

## Methods

**Materials and Structure.**

Synthesis of BAPI Single Crystals: PbO powder (2232 mg, 10 mmol) was dissolved in a mixed acid solution of hydroiodic acid (HI, 10.0 mL, 76 mmol) and hypophosphorous acid ($H_3PO_2$, 1.7 mL, 15.5 mmol) under constant magnetic stirring to obtain a clear, bright-yellow $PbI_2$ precursor solution. Meanwhile, Butylamine (BA, 924 μL, 10 mmol) was neutralized with excess HI (5 mL, 38 mmol) in an ice bath to prepare the BAI precursor solution. The resulting BAI solution was then added dropwise into the preheated $PbI_2$ solution maintained at 100 °C in an oil bath. An immediate orange precipitate appeared, which gradually dissolved upon continuous stirring, forming a transparent, bright-yellow precursor solution. The solution was subsequently cooled to room temperature and left undisturbed for approximately 2 hours, resulting in the spontaneous formation of high-quality BAPI single crystals.

The structure fabrication followed the steps below (**Fig. S1**). First, a 50 nm Ag film was deposited onto a $Si/SiO_2$ substrate by magnetron sputtering, followed by a 5 nm $SiO_2$ thin layer using the same method. The BAPI single crystal was then mechanically exfoliated and transferred onto the $SiO_2$ film, providing BAPI flakes with various thicknesses. Subsequently, a PMMA solution dissolved in toluene was spin-coated uniformly onto the substrate. After allowing the film to stabilize, a 40 nm Ag layer was deposited on top of the PMMA by thermal evaporation.

**Optical Characterizations.**

Absorption spectra were acquired using a home-built micro transmission/absorption spectrometer. Broadband illumination was provided by a halogen lamp (SLS201L, Thorlabs) and focused onto the sample through a high-numerical-aperture objective lens (Olympus, × 50, NA = 0.8). The transmitted light was subsequently collected by a second objective, collimated, and coupled into a fiber optic cable before being analyzed by a liquid nitrogen-cooled spectrometer (HRS-300, Princeton Instruments). Photoluminescence spectra were measured using a similar instrumental configuration, where the halogen lamp was replaced with a 405 nm continuous-wave laser.

Angle-resolved reflectance spectra were obtained using a custom-built 4 f Fourier imaging system. Angular information was collected through a 100 × objective lens (Olympus, × 100, NA = 0.9), and an angular resolution of ± 64° was obtained through slit splitting in the Fourier phase plane. The signal was collected through a liquid nitrogen-cooled spectrometer (HRS-300, Princeton Instruments), and analyzed with 600 nm gratings to obtain angle-resolved reflectance spectra. The white light source for the reflectance spectra comes from a tungsten halogen lamp source (SLS201L, Thorlabs).

The schematic of the transient reflectance microscopy setup is shown in SI **Fig. S8**. A train of 800 nm pulses was generated using a Coherent Astrella regenerative amplifier (80 fs, 1 kHz, 2.5 mJ/pulse), seeded by a Coherent Vitara-s oscillator (35 fs, 80 MHz). These pulses were used to pump an optical parametric amplifier (OPA; Coherent OperA Solo) to generate both pump and probe beams. In our experiments, the pump beam at 400 nm was produced by frequency-doubling the 800 nm pulses using a β-barium borate (BBO) crystal, while the probe pulses were generated by the OPA. A controllable time delay between the pump and probe pulses was introduced by placing a motorized translation stage (DDS220/M, Thorlabs) in the probe path. Both pump and probe beams were directed onto the sample through a 60× reflective objective (Olympus UPLFLN, × 60, NA = 0.9). The probe beam reflected from the sample was collected by the same objective and relayed by a 200 mm tube lens onto a CMOS detector (PL-D755MU-T, Pixelink) for imaging.

**Theory.**

The dynamics of polaritons under non-resonant excitation can be described by a driven-dissipative Gross-Pitaevskii (GP) equation:

$$i\hbar \frac{\partial \psi(\boldsymbol{r},t)}{\partial t} = \left[ -\frac{\hbar^2}{2m_{eff}}\nabla^2 - i\hbar\frac{\gamma_c}{2} + g_c|\psi(\boldsymbol{r},t)|^2 + (g_r + i\hbar\frac{R}{2})n_A(\boldsymbol{r},t) + g_r n_I(\boldsymbol{r},t) \right]\psi(\boldsymbol{r},t).$$

Here, $\psi(\boldsymbol{r},t)$ is the polariton field, $m_{eff} = 3\times 10^{-5} m_e$ ($m_e$ is the free electron mass) is the effective mass of the lower polariton branch under parabolic approximation, $\gamma_c = 3$ ps$^{-1}$ is the polariton loss rate, $g_c = 0.1$ μeV·μm² represents the polariton-polariton interaction, $g_r = 0.2$

μeV·μm² represents the polariton-reservoir interaction with both the active reservoir $n_A$ and inactive reservoir $n_I$, and $R = 2$ ps⁻¹·μm² is the condensation rate. The density of the active reservoir satisfies

$$\frac{\partial n_A(\mathbf{r},t)}{\partial t} = \tau n_I(\mathbf{r},t) - \gamma_A n_A(\mathbf{r},t) - R|\psi(\mathbf{r},t)|^2 n_A(\mathbf{r},t).$$

The inactive reservoir contains hot excitons excited directly by the external non-resonant pump $P(\mathbf{r},t)$ and obeys the following equation of motion:

$$\frac{\partial n_I(\mathbf{r},t)}{\partial t} = -\tau n_I(\mathbf{r},t) - \gamma_I n_I(\mathbf{r},t) + P(\mathbf{r},t).$$

Here, $\gamma_A = 0.01$ ps⁻¹ and $\gamma_I = 0.01$ ps⁻¹ are the loss rates of the active and inactive reservoirs, respectively, and $\tau = 0.1$ ps⁻¹ is the relaxation rate of the inactive reservoir to the active reservoir. The coupled equations can be solved efficiently by a numerical solver[42]. The time evolutions in **Figs. 3a-3c** are extracted at specific energies that correspond to the targeted wavevectors.

# Reference

1. Hopfield, J. J. Theory of the Contribution of Excitons to the Complex Dielectric Constant of Crystals. *Phys. Rev.* **112**, 1555–1567 (1958).
2. Weisbuch, C., Nishioka, M., Ishikawa, A. & Arakawa, Y. Observation of the coupled exciton-photon mode splitting in a semiconductor quantum microcavity. *Phys. Rev. Lett.* **69**, 3314–3317 (1992).
3. Kasprzak, J. *et al.* Bose-Einstein condensation of exciton polaritons. *Nature* **443**, 409–414 (2006).
4. Gao, T. *et al.* Observation of non-Hermitian degeneracies in a chaotic exciton-polariton billiard. *Nature* **526**, 554–558 (2015).
5. Peng, K. *et al.* Topological valley Hall polariton condensation. *Nat. Nanotechnol.* **19**, 1283–1289 (2024).
6. Trypogeorgos, D. *et al.* Emerging supersolidity in photonic-crystal polariton condensates. *Nature* **639**, 337–341 (2025).
7. Li, H. *et al.* All-optical temporal logic gates in localized exciton polaritons. *Nat. Photon.* **18**, 864–869 (2024).
8. Shi, Y. *et al.* Coherent optical spin Hall transport for polaritonics at room temperature. *Nat. Mater.* **24**, 56–62 (2025).
9. Kędziora, M. *et al.* Predesigned perovskite crystal waveguides for room-temperature exciton-polariton condensation and edge lasing. *Nat. Mater.* **23**, 1515–1522 (2024).
10. Wu, X. *et al.* Exciton polariton condensation from bound states in the continuum at room temperature. *Nat Commun* **15**, 3345 (2024).
11. Song, J. *et al.* Room-temperature continuous-wave pumped exciton polariton condensation in a perovskite microcavity. *Sci. Adv.* **11**, eadr1652 (2025).
12. Mavrotsoupakis, E. G. *et al.* Unveiling asymmetric topological photonic states in anisotropic 2D perovskite microcavities. *Light Sci Appl* **14**, 207 (2025).
13. Barrat, J. *et al.* Qubit analog with polariton superfluid in an annular trap. *Sci. Adv.* **10**, eado4042 (2024).
14. Kavokin, A. *et al.* Polariton condensates for classical and quantum computing. *Nat Rev Phys* **4**, 435–451 (2022).
15. Klembt, S. *et al.* Exciton-polariton topological insulator. *Nature* **562**, 552–556 (2018).
16. Balasubrahmaniyam, M. *et al.* From enhanced diffusion to ultrafast ballistic motion of hybrid light-matter excitations. *Nat. Mater.* **22**, 338–344 (2023).
17. Pandya, R. *et al.* Microcavity-like exciton-polaritons can be the primary photoexcitation in bare organic semiconductors. *Nat Commun* **12**, 6519 (2021).
18. Xu, D. *et al.* Ultrafast imaging of polariton propagation and interactions. *Nat Commun* **14**, 3881 (2023).
19. Xie, X. *et al.* 2D material exciton-polariton transport on 2D photonic crystals. *Sci. Adv.* **11**, eads0231 (2025).
20. Wertz, E. *et al.* Propagation and Amplification Dynamics of 1D Polariton Condensates. *Phys. Rev. Lett.* **109**, 216404 (2012).
21. Krupp, N., Groenhof, G. & Vendrell, O. Quantum dynamics simulation of exciton-polariton transport. *Nat Commun* **16**, 5431 (2025).

22. Liu, Y.-C. *et al.* Coherent Polariton Dynamics in Coupled Highly Dissipative Cavities. *Phys. Rev. Lett.* **112**, 213602 (2014).
23. Anantharaman, S. B. *et al.* Dynamics of self-hybridized exciton-polaritons in 2D halide perovskites. *Light Sci Appl* **13**, 1 (2024).
24. Knorr, M. *et al.* Intersubband Polariton-Polariton Scattering in a Dispersive Microcavity. *Phys. Rev. Lett.* **128**, 247401 (2022).
25. Michetti, P. & La Rocca, G. C. Polariton-polariton scattering in organic microcavities at high excitation densities. *Phys. Rev. B* **82**, 115327 (2010).
26. Laitz, M. *et al.* Uncovering temperature-dependent exciton-polariton relaxation mechanisms in hybrid organic-inorganic perovskites. *Nat Commun* **14**, 2426 (2023).
27. Tassone, F., Piermarocchi, C., Savona, V., Quattropani, A. & Schwendimann, P. Bottleneck effects in the relaxation and photoluminescence of microcavity polaritons. *Phys. Rev. B* **56**, 7554–7563 (1997).
28. Perrin, M., Senellart, P., Lemaître, A. & Bloch, J. Polariton relaxation in semiconductor microcavities: Efficiency of electron-polariton scattering. *Phys. Rev. B* **72**, 075340 (2005).
29. Zhao, J. *et al.* Exciton polariton interactions in Van der Waals superlattices at room temperature. *Nat Commun* **14**, 1512 (2023).
30. Claude, F. *et al.* Observation of the diffusive Nambu-Goldstone mode of a non-equilibrium phase transition. *Nat. Phys.* **21**, 924–930 (2025).
31. Ballarini, D. *et al.* Directional Goldstone waves in polariton condensates close to equilibrium. *Nat Commun* **11**, 217 (2020).
32. Su, R. *et al.* Perovskite semiconductors for room-temperature exciton-polaritonics. *Nat. Mater.* **20**, 1315–1324 (2021).
33. Bourelle, S. A. *et al.* Optical control of exciton spin dynamics in layered metal halide perovskites via polaronic state formation. *Nat Commun* **13**, 3320 (2022).
34. Mavrotsoupakis, E. G. *et al.* Unveiling asymmetric topological photonic states in anisotropic 2D perovskite microcavities. *Light Sci Appl* **14**, 207 (2025).
35. Jin, F. *et al.* Observation of perovskite topological valley exciton-polaritons at room temperature. *Nat Commun* **15**, 10563 (2024).
36. Tao, R. *et al.* Halide perovskites enable polaritonic XY spin Hamiltonian at room temperature. *Nat. Mater.* **21**, 761–766 (2022).
37. Li, Y. *et al.* Manipulating polariton condensates by Rashba-Dresselhaus coupling at room temperature. *Nat Commun* **13**, 3785 (2022).
38. Zhang, M. *et al.* Two-dimensional organic-inorganic hybrid perovskite quantum-well nanowires enabled by directional noncovalent intermolecular interactions. *Nat Commun* **16**, 2997 (2025).
39. Gong, Y. *et al.* Boosting exciton mobility approaching Mott-Ioffe-Regel limit in Ruddlesden-Popper perovskites by anchoring the organic cation. *Nat Commun* **15**, 1893 (2024).
40. Yue, S. *et al.* High ambipolar mobility in cubic boron arsenide revealed by transient reflectivity microscopy. *Science* **377**, 433–436 (2022).
41. Ma, X. *et al.* Realization of all-optical vortex switching in exciton-polariton condensates. *Nat Commun* **11**, 897 (2020).
42. Wingenbach, J. *et al.* PHOENIX - Paderborn highly optimized and energy efficient solver for two-dimensional nonlinear Schrödinger equations with integrated extensions. *Comput.*

*Phys. Commun.* **315**, 109689 (2025).

## Acknowledgements

The authors acknowledge the financial support from the National Key Research and Development Program of China (2023YFA1507002 and 2024YFA1208203), the Strategic Priority Research Program of the Chinese Academy of Sciences (XDB0770000), the National Science Foundation for Distinguished Young Scholars of China (No. 22325301), the Natural Science Foundation of China (22173025, 22073022 and U23A2076).

## Author information

Author contributions: X.L. and Q.Z. led the project. X.L., Q.Z. and Y.X. conceived the idea. X.L., Q.Z. and Y.X. designed experiments. Y.L. produced the samples. Y.X., Z.Z. and X.B. performed physical characterization. Y.X., T.L., Y.Z. and X.Z. conducted the angle-resolved spectra experiments. Y.X. conducted the TRM experiments. X.M. and S.S. performed the numerical calculations. Y.X., Y.G., Y.Z., X.Z., S.Y., W.D., J.F. and R.S. led the experimental analysis. X.L., Q.Z., X.M., Y.X., R.S. and S.S. prepared the manuscript. All authors discussed the results and revised the manuscript.

## Ethics declarations

Competing interests

The authors declare no competing interests.

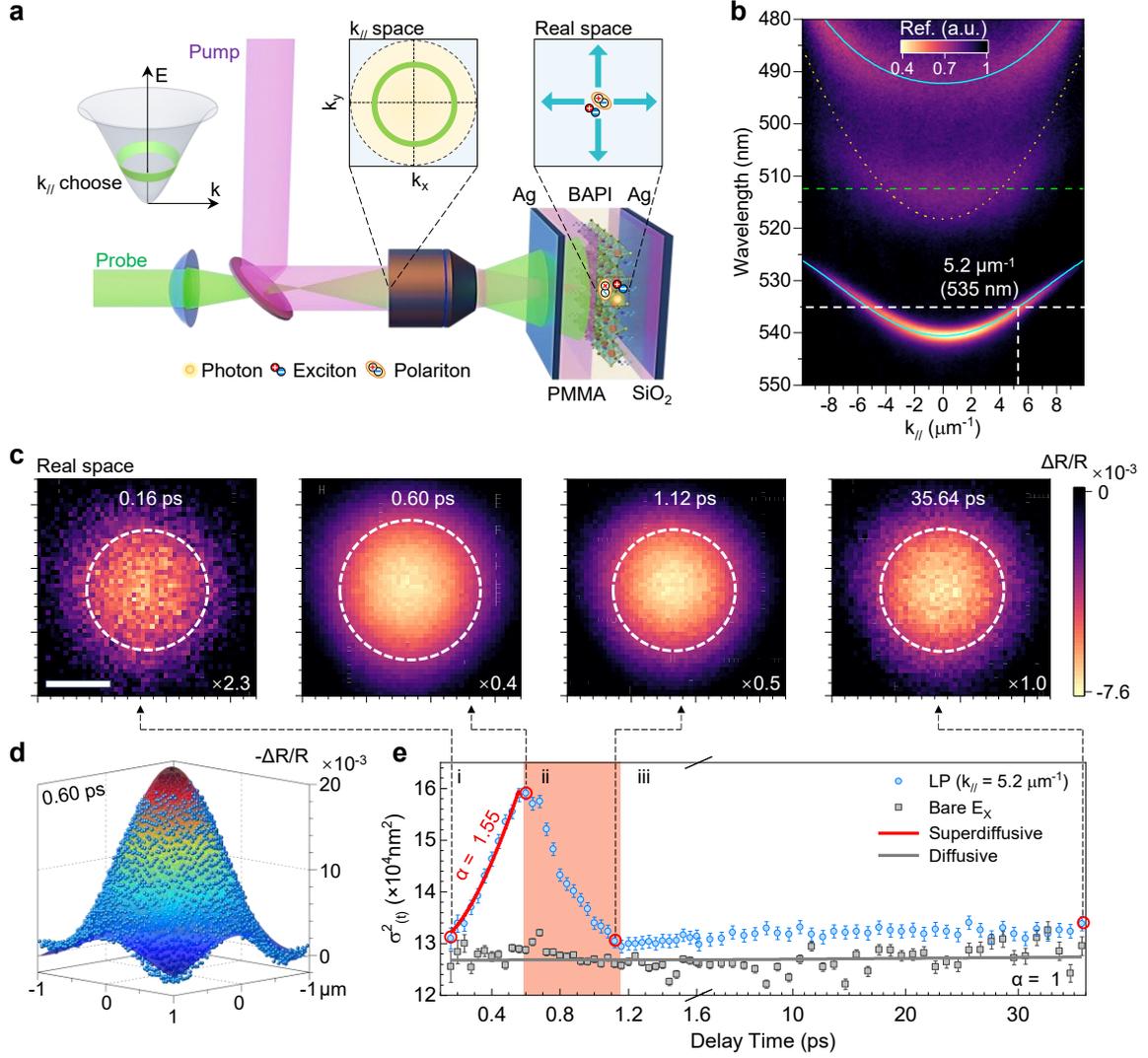

**Figure 1. Investigation of EP transport dynamics via $k_{//}$-resolved TRM.** (a) Schematic of $k_{//}$-resolved TRM and the microcavity structure. (b) Angle-resolved reflection spectrum of BAPI in the FP cavity. The white dashed line corresponds to the $k_{//}$ value (and corresponding wavelength) for subsequent measurements. (c) 2D ΔR/R distribution at characteristic times of 0.16, 0.6, 1.12, and 35.64 ps. The white scale bar corresponds to 500 nm. (d) 3D plot of the spatially resolved ΔR/R intensity at 0.6 ps. The rainbow-colored surface is the corresponding surface obtained through 2D Gaussian fitting. (e) Temporal evolution of the variance, $\sigma^2(t)$, extracted from 2D Gaussian fitting of ΔR/R, across the full-time range of all measurements. Blue dots represent polariton transport measured at $k_{//} = 5.2$ μm$^{-1}$, while gray dots correspond to exciton transport in the absence of the FP cavity, probed at bare exciton absorption peak.

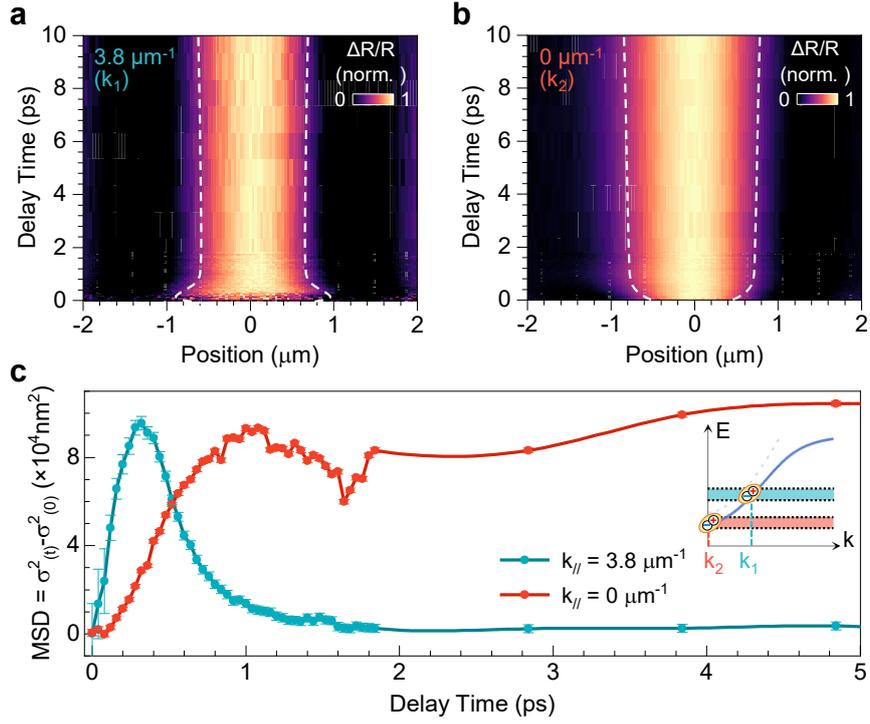

**Figure 2. Temporal Evolution of EP Population at Different $k_{//}$.** Time evolution of the normalized $\Delta R/R$ at (a) $k_{//} = 3.8$ μm$^{-1}$ and (b) $k_{//} = 0$ μm$^{-1}$, used to characterize the polariton population distribution. (c) Temporal evolution of the fitted spatial variance $\sigma^2(t)$ of the population distribution. The salmon curve corresponds to (a) and the cyan curve to (b). The inset illustrates the corresponding EP dispersion and the probe regions.

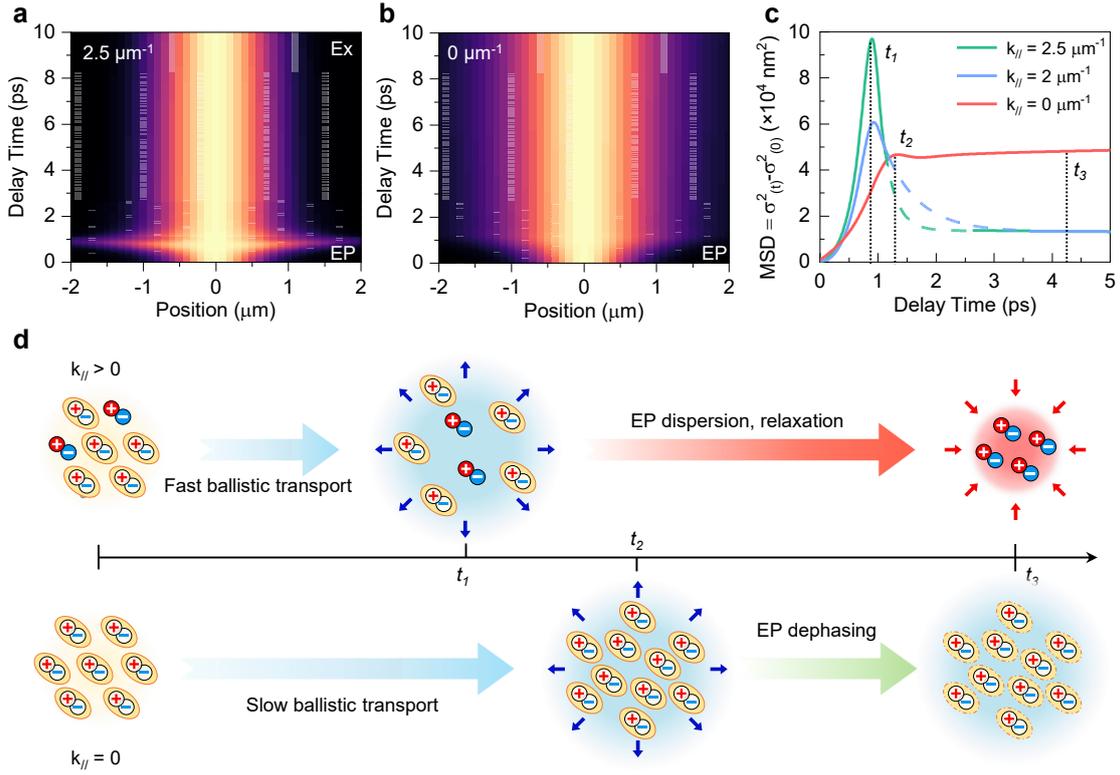

**Figure 3. Numerical simulations of spatiotemporal dynamics of $k_{//}$-dependent EPs.** Time evolution of the EP (or exciton) population distribution at (a) $k_{//} = 2.5$ μm$^{-1}$ and (b) $k_{//} = 0$ μm$^{-1}$. (c) Temporal evolution of the fitted spatial variance $\sigma^2(t)$ of the population distribution at $k_{//} = 2.5$, 2 and 0 μm$^{-1}$. (d) Phenomenological illustration of the evolution of EPs and excitons at $k_{//} > 0$ and $k_{//} = 0$.

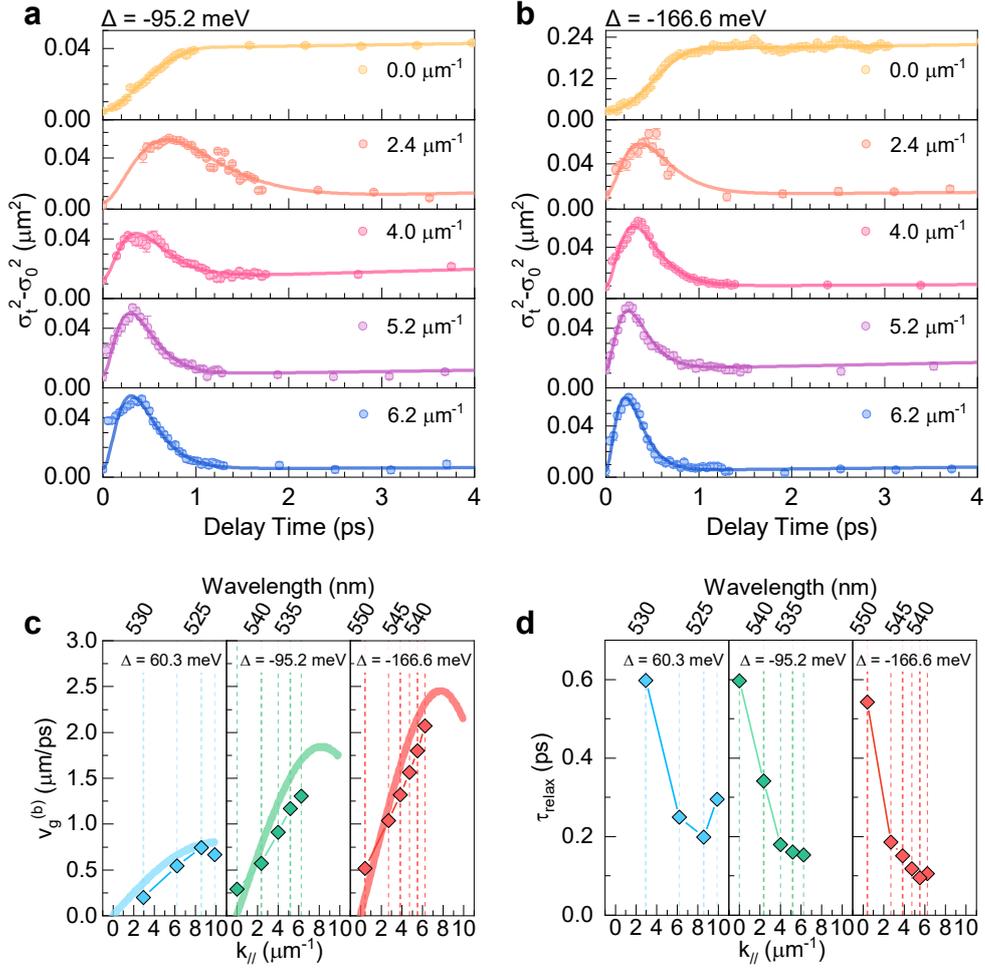

**Figure 4. Manipulation of EP Transport Dynamics by Detuning.** (a) and (b) show the evolution of the MSD of EPs over time for different $k_{//}$ at $\Delta$ = -95.2 meV and $\Delta$ = -166.6 meV, respectively. (c) and (d) present the extracted group velocities of ballistic transport $v_g^{(b)}$ and relaxation times $\tau_{relax}$ of EPs, obtained by analyzing the transport dynamics at $\Delta$ = 60.3, -95.2, and -166.6 meV.